\documentstyle[11pt,paspconf]{article}
\begin{document}
\title{A Fokker-Planck Model of Rotating Stellar Clusters}
\author{John Girash}
\affil{Center for Astrophysics, 60 Garden St., Cambridge MA 02138 USA}

\begin{abstract}
 We have developed a two-dimensional orbit averaged Fok\-ker-Planck model of
 stellar clusters which expands on spherically symmetric
 one-dimensional models to include rotation and ellipticity.
 Physical effects such as collisions, finite stellar lifetimes
 and bar formation ({\it i.e.,} a non-axisymmetric component of the potential)
 can also be included. The first use of the model is to study the evolution
 of dense clusters ($\rho_\circ\simeq10^7M_\odot/{\rm pc}^3$) that may be
 expected to have existed at the centres of newly-forming galaxies, with the
 goal of verifying that angular momentum can be removed from the core of the
 cluster quickly enough so that rotation no longer prevents the formation of
 a massive ($\sim10^2 M_\odot$) object. This could act as the seed black
 hole for the formation of an AGN.

\end{abstract}

\keywords{stars: stellar dynamics, galaxies: internal motions, bars}
\section{Introduction}

Quinlan \& Shapiro (1989) developed a Fokker-Planck model of a spherically
symmetric, dense cluster of compact stars, and found ``rapid buildup of
massive black holes in the cluster core resulting from successive binary
mergers and mass segregation.''  Subsequently, they studied clusters of
solar-mass stars and found ``it is remarkably easy for massive stars to
form through multiple stellar mergers in dense galactic nuclei''
(Quinlan \& Shapiro, 1990; hereafter QS).

The goal of this project is to generalise the QS approach to two dimensions
in order to answer the question: What about rotation?  The hoped-for
answer is that mergers and mass segregation can still produce a massive object
(perhaps $\simeq10^{2-3}M_\odot$) in the core of the cluster, which could 
then undergo growth via accretion to reach
supermassive size ($\sim10^{6-8}M_\odot$) within a Hubble time ({\it 
e.g.\/}, David {\it et al.\/}, 1987) and be seen as an AGN.

\section{The Orbit-averaged Fokker-Planck equation}

Because of the non-spherical nature of our model, it is not possible to use
$(E,J_z)$ as our integrals of the motion as previous two-dimensional work
has done (Cohn 1979).  Instead we use radial ($I_1$) and tangential ($I_2$)
action variables, which are adiabatic invariants as the potential evolves.
Orbit-averaging is then truly a simple average of quantities over the $2\pi$
change in (an) angle variable. The orbit-averaged Fokker-Planck equation
then takes a particularly simple form:
\begin{displaymath}
{\partial \over\partial t}
f_n=-{\partial \over\partial I_j}
[f_n\left<\Delta I_j\right>]+{1\over2}
{\partial^2\over\partial I_i\partial I_j}[f_n\left<\Delta I_i\Delta I_j\right>]
-L_n+G_n-B_n+R_n
\end{displaymath}
where $f_n(I_1,I_2,t)$ are the distribution functions for stars of mass-type
$n$ and $\left<.\right>$ are the drift and diffusion coefficients. $L_n, G_n$
are ad-hoc terms to account for losses and gains due to stellar mergers, while
$B_n$ and $R_n$ are the analogous paramters for stellar evolution.  
Summation over repeated dummy indices $i$ and $j$ is implied.

The coefficient calculation (partially derived by Tremaine and
Weinberg, 1984, and by Van Vleck, 1926, in different contexts)
involves expanding the potential in action space and summing over
the entire distribution $f=\sum_nf_n$:
\begin{displaymath}
\left<\Delta I_j\right>=-2\pi^3\int dI_1dI_2
\left(\sum_{k=1}^2\ell_k{\partial f\over\partial I_k}\right)
\sum_{\ell_1\ell_2\ell_3\pm}\ell_j~|\Psi_{\ell_1\ell_2\ell_3}|^2~
\delta(\ell_1\Omega_1+\ell_2\Omega_2\pm\ell_3\Omega_b),
\nonumber\end{displaymath}
\begin{displaymath}
\left<\Delta I_i\Delta I_j\right>=-4\pi^3\int dI_1dI_2~f
\sum_{\ell_1\ell_2\ell_3\pm}\ell_i\ell_j~|\Psi_{\ell_1\ell_2\ell_3}|^2~
\delta(\ell_1\Omega_1+\ell_2\Omega_2\pm\ell_3\Omega_b).
\nonumber\end{displaymath}
In the above, $\Omega$ are the orbital frequencies, {\it i.e.,}
$\Omega_1={\partial E/\partial I_1}$, and the $\ell$ are integers labelling 
different coefficients $\Psi_{\ell_1\ell_2\ell_3}$ in the expansion of the 
potential in action space.  Subscript $b$ signifies ``bar'', but 
should be understood to represent any individual element of the potential, 
{\it e.g.,} that of one particular ``field'' star.

The Fokker-Planck approximation remains valid as long as number 
$N_n=\int dI_1dI_2f_n\gg1$, and requires that the time step $\Delta t$ satisfy
$t_{dyn}\ll\Delta t\ll t_{r}$. We have dynamical time $t_{dyn}<\simeq10^4 y$,
and relaxation time $t_{r}>10^6 y$ or more.

\section{The Potential}

We must iteratively solve for the self-consistent potential $\Phi$ after
each Fokker-Planck time step.  To allow for non-sphericity, we first
calculate the ellipticity $\epsilon$ from the overall net rotational
velocity 
using the virial theorem (Binney \& Tremaine, 1987). 
To make the problem tractable, it is assumed that isodensity surfaces are
all of this common ellipticity.  The density is thus expressed in terms of the
homeoidal radial coordinate $m^2\equiv R^2+{z^2\over1+\epsilon}$:
\begin{displaymath}
\rho(m^2)={1\over4\pi^2m}\int{dI_1 dI_2\over I_2}~\Omega_1\sum_nM_nf_n
\end{displaymath}
in which $M_n$ denotes the mass of star type $n$, and
from which the new potential run can be calculated.
This procedure is iterated until  convergence is achieved, typically to
$\sim1\%$, and accelerated by the Aitken ``$\Delta^2$'' process
(Henrici, 1964).  A final check of conservation of overall energy
is also made.
The $m^2$ grid is variable with each time step, with regions of larger
$d\Phi/dm^2$ and $d\rho/dm^2$ being given a greater density of grid points.

The assumption of homeiodal isodensity surfaces, along with $f_n$,
determines the potential $\Phi$.  The required integral, however, is too
involved to be performed each time the knowledge of potential is needed in
the Fokker-Planck coefficient calculation.  After testing interpolation
schemes and analytic approximations, only the Clutton-Brock self-consistent
field method (Hernquist and Ostriker, 1992) proved adequate in both accuracy
and speed. The field method is tested at each timestep, and if sufficient
accuracy cannot be achieved with a reasonable number of expansion terms, the
code falls back on direct integration.


\section{The Bar}

Although some angular momentum is expected to be transported outwards via
the effects of shear between the higher-$\Omega$ inner regions and the
lower-$\Omega$ outer regions, we are appealing to a bar-like perturbation in 
the potential to do most of the angular momentum transfer.  So, should the
cluster be found to have become unstable to formation of a stellar bar 
({\it i.e.}, be found to satisfy a $T_{rot}/|W|$-style instability criterion
such as that of Christodoulou {\it et al.}, 1995), a non-axisymmetric
component is incorporated into the potential.  Combes \& Sanders (1981) found
that bars form over 1--2 rotational periods, and last for many more ($>10$).
This has led us to build the nonaxisymmetric potential from a fraction
($\sim10\%$) of the $1~M_\odot$ population that is forced to
orbit in ``locked step'', sharing a common orbital frequency and phase.
We assume there is no transfer of stars from bar to field or vice versa. 
Combes \& Elmegreen (1993) show that the bar frequency is a compromise of
the orbital frequencies of its component stars, so our $\Omega_b$
is set by
conserving either the total energy or total angular momentum of those
stars.  Allowing the bar distribution to evolve like the
field stars 
avoids any problem of inserting  the bar binding energy by hand.

\begin{figure} [tb]

\vspace{2.4in}
\includegraphics{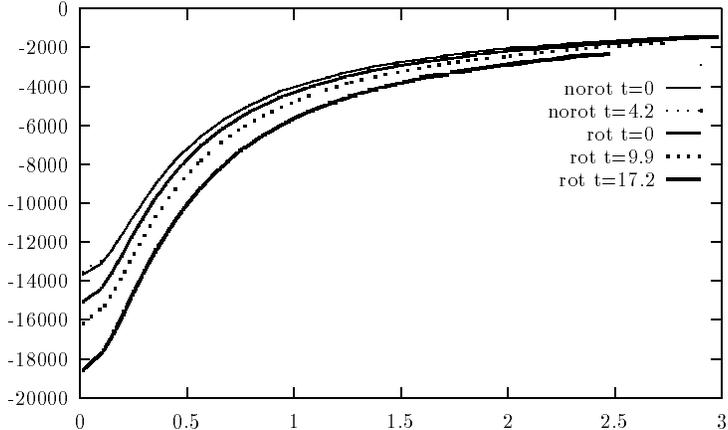}

\caption{Gravitational potential $\Phi$ 
in units of km/s as a function of equatorial distance $r$ [pc] for an
$N=10^6$ Plummer sphere and a modified cluster of $1~M_\odot$ stars with
core radius 0.3 pc. It is probably wise not to put too much faith
in the $t=17.2$ line, as by that time the energy error $\Delta E/E$ had
reached 100\%. At $t=9.9$, $\Delta E/E\simeq10\%$.  The time unit is
$0.98\times10^6$ yrs.  $t_r(r=0)\simeq9\times10^6y$.}

\end{figure}

\section{Results}

It is standard in this game to start with a Plummer sphere distribution.
To introduce rotation, we start with the rotation parameter
$\lambda\equiv J_{rot}|W_{grav}|^{1\over2}/GM_{tot}^{2.5}$,
and ``shift'' the distribution of $f(J)$ so that total $J$ matches desired
$J_{rot}$ while conserving number, so that the new $f_\lambda(J)=0$ for $J<$
some $J_{min}$, and $f_\lambda=const\times f$ for $J>J_{min}$.
A typical $\lambda$ value from cosmological tidal torques is $\sim$0.05
(Barnes \& Efstathiou, 1987).  While this prescription does
produce an overall rotation, it does so at the ``expense'' of the tangential 
velocity dispersion $\sigma_t^2$, {\it i.e.}, the fraction
of radial orbits is enhanced and the tangential component of the kinetic
energy is actually decreased.  From a computational ``proof-of-concept''
point of view, however, this increase in radial orbits is an advantage as it
plays into the density dependence of the gravitational relaxation.
\begin{figure} [tb]

\vspace{2.4in}
\includegraphics{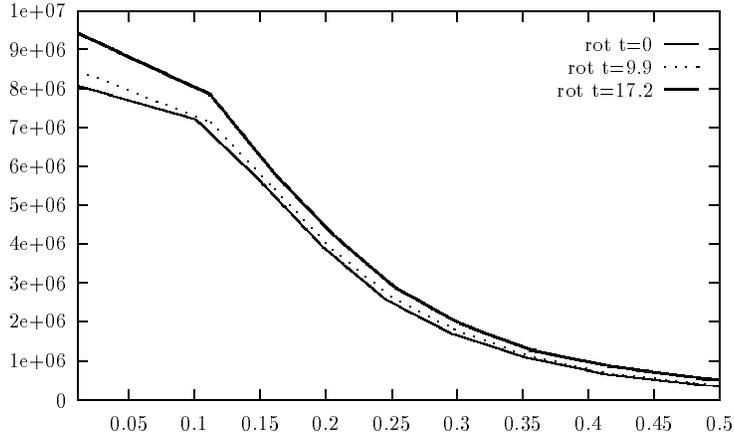}

\caption{Density [$M_\odot/pc^3$] of the inner $0.5$ pc of the modified
cluster.}

\end{figure}

Figure 1 details the evolution of the potential run for $N=10^6$ clusters 
of $1~M_\odot$ stars.  The initally Plummer-like cluster shows no
change over half of a central relaxation time ($\simeq9\times10^6$ yrs).
Unfortunately there was not time for a longer run prior to this meeting, but
this demonstrates the stability of the staggered distribution/potential
updating scheme.  The modified cluster was run on a faster machine, so
there was time to evolve it for $\sim2$ central relaxation times.  Note how
its potential starts out deeper and deepens more rapidly than that of the
Plummer cluster, as would be expected when radial orbits are more dominant
and stars spend more time in the denser core where two-body relaxation
has a greater effect.  The corresponding densities are shown in Figure 2;
the breaks in $\rho(r)$ are caused by insufficient resolution in the
innermost region.\footnotemark
\footnotetext{Further similar results were presented at the meeting, but
in light of refinements made since, I will use the remainder of this space
to show the equivalent newer results.}

\section{Conclusions and ``To-do's''}
We have demonstrated the stability of this method for modeling the evolution
of a dense star cluster using the two-dimensional Fokker-Planck equation,
but work needs to be done to balance keeping $\Delta E/E$ small while
maintaining a reasonable time step $\Delta t$. Ways to accomplish this
include allowing a variable $\Delta t$ when $(\Delta E/E)/\Delta t$ becomes
too large (which introduces its own issues of energy error; these may be
alleviated using the forced-time-symmetry technique described elsewhere in 
these proceedings by Piet Hut), or simply by throwing more cycles at the problem.
We should also weight the small-$r$ grid more.  On the physical side of the
problem, the initial conditions need to be chosen for higher $v_{rms}$ (a
criticism of QS was that they chose initial conditions favouring massive
object formation, but we should at least reproduce their result as a test
case), and also not to cool $\sigma_t^2$ when rotation is desired.  The
bar can then be ``turned on'', as can the effects of stellar mergers and evolution.

\begin{figure} [tb]

\vspace{2.4in}
\includegraphics{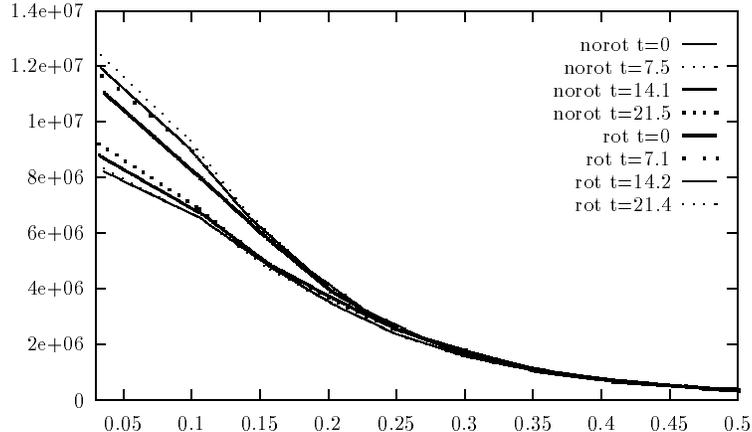}

\caption{New density runs for similar clusters as in Figure 2.}

\end{figure}

\section{Post-Denouement}

In the time between the Kingston meeting and the proceedings deadline, some
progress towards these goals has been made.  The energy error has been found
to be largely dependent upon the differencing scheme used for the
distribution functions $f_n(I_1,I_2)$. The Chang-Cooper scheme, which in one
dimension guarantees that a Fokker-Planck distribution will remain positive
for any $\Delta t$, does not generalise to 2-D (essentially
because the calculus of extrema becomes more complicated).  For the
meeting, I used a Chang-Cooper-based method that overcompensated
(treating all derivatives as full) but which in later tests proved
unsatisfactory, as did undercompensating (using partials).  
Figure 3 shows the results of
a simple fixed differencing weighted 90\% towards the forward side,
which allowed the $1~M_\odot$ case to run for more than two central
relaxation times ({\it i.e.}, over four times as long as before) before
$\Delta E/E$ reached $\sim10\%$.  One can now see the beginnings of density
increase in the core.  Also, the break in $\rho(r)$ at $r\simeq0.1$ pc has
been almost eliminated by the use of a better-optimised $m^2$ grid.

\begin{figure} [tb]

\vspace{2.4in}
\includegraphics{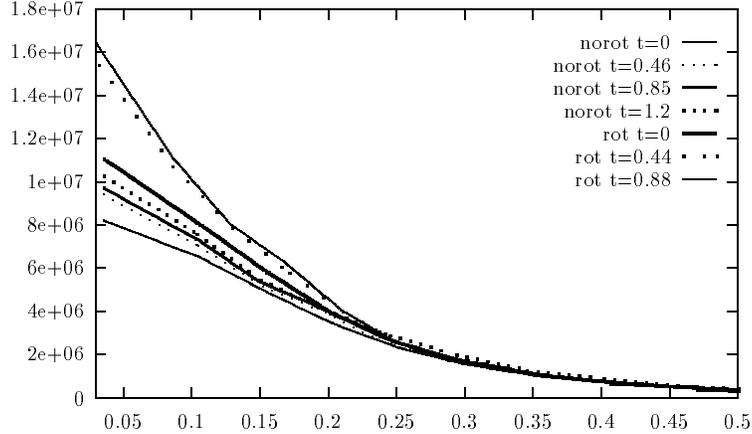}

\caption{$\rho(r)$ for clusters of 60\%
$1M_\odot$, 30\% $2M_\odot$, 10\% $4M_\odot$ by mass.}

\end{figure}

The results for clusters of the same overall mass, but with a mix of 1-, 2-
and 4-$M_\odot$ stars is shown in Figure 4.  In this case, $\Delta E/E$
reached $\sim20\%$ after just over one core relaxation time of $\sim0.75$
had passed. 
Here we see that both clusters have
larger rates of central density increase than the equivalent all-$1~M_\odot$
clusters, and that the relative increase of the rate for the cluster with
enhanced radial orbits (quixotically labelled ``rot'' as described in \S5)
relative to the Plummer-like one (``norot'') is much quicker than in the
$1~M_\odot$ case. Both of these differences are as expected when the
evolution is dominated by two-body gravitational relaxation.

\acknowledgments

In addition to the insightful support of my advisor George Field, I would
like to acknowledge Bill Press, Charles Gammie and Ue-Li Pen for their
advice, and NCSA for cycles.  Special thanks go to David Clarke, for
patience while editing.

\end{document}